\documentclass[useAMS,usenatbib]{mn2e}
\input psfig.tex

\title[RzCS 052 and other high redshift clusters]{Scaling relations of the colour--detected
cluster RzCS 052 at $\mathbf{z=1.016}$ and some other high redshift clusters}
\author[Andreon et al.]{S. Andreon,$^1$\thanks{stefano.andreon@brera.inaf.it},
R. De Propris$^2$, E. Puddu$^3$, L. Giordano$^{1,4}$, H. Quintana$^{5}$ \\
$^1$INAF--Osservatorio Astronomico di Brera, via Brera 28, 20121, Milano, Italy \\
$^2$Cerro Tololo Inter--American Observatory, La Serena, Chile\\
$^3$INAF--Osservatorio Astronomico di Capodimonte, salita Moiariello 16, 80131 Napoli, Italy\\
$^4$Universit\`a degli Studi di Milano Bicocca, Piazza della Scienza 3, 20126 Milano, Italy\\
$^5$Departamento de Astronom\'\i a y Astrof\'\i sica, Pontificia Universidad Cat\'olica de Chile, Santiago, Chile\\
}
\date{Accepted ... Received ...}
\pagerange{\pageref{firstpage}--\pageref{lastpage}}
\pubyear{2007}
\begin{document}
\maketitle

\label{firstpage}

\begin{abstract}

We report on the discovery of the $z=1.016$ cluster RzCS 052 using a modified
red sequence method, followup spectroscopy and X-ray imaging.  This cluster
has a velocity dispersion of $710 \pm 150$ km s$^{-1}$, a virial
mass of $4.0 \times 10^{14}\ M_{\odot}$ (based on 21 spectroscopically
confirmed members) and an X-ray luminosity of $(0.68 \pm 0.47) \times 10^{44}$
ergs s$^{-1}$ in the [1-4] keV band. This optically selected cluster appears to
be of richness  class 3 and to follow the known $L_X - \sigma_v$ relation for
high redshift X-ray selected clusters. Using these data, we find that the halo
occupation number for this cluster is only marginally consistent with what
expected assuming a self-similar evolution of cluster scaling relations,
suggesting perhaps a break of them at $z\sim1$. We also rule out
a strong galaxy merging activity between $z=1$ and today. Finally, we
present a Bayesian
approach to measuring cluster velocity dispersions and X-ray luminosities in
the presence of a background: we critically reanalyze recent claims for X-ray
underluminous clusters using these techniques and find that the clusters can be
accommodated within the existing $L_X - \sigma_v$ relation.

\end{abstract}


\section{Introduction}

Clusters of galaxies are not only a powerful tool to study galaxy evolution
but can also be used to constrain cosmological parameters, resolving several
parameter degeneracies (e.g.,  Allen et al. 2004;  Albrecht et al. 2006). 
In particular, clusters at high redshifts ($z > 1$), of which only a handful 
are currently known, provide the greatest leverage in determining the nature of
the acceleration constant (e.g., Rapetti 2007). These determinations,
however, rely on an accurate estimate of the cluster mass, whose uncertainty
is arguably the dominant contributor to the error budget in deriving cosmological
parameters from cluster statistics (Henry 2004; Albrecht et al. 2006).

Ideally, one wish to apply the virial theorem to get a direct measurement
of cluster masses. In fact, the dark matter velocity dispersion is an
extremely good tracer of the halo masses in all simulations 
(Evrard et al. 2007), and galaxies
are nearly unbiased velocity tracer (Evrard et al. 2007 and references
therein; Rines Diaferio \& Natarajan 2007), in good agreement 
previous works (Biviano et al. 2006, Tormen et al. 1997).
The measurement of the cluster velocity dispersion
requires a large number of radial velocities, which
are observationally expensive to obtain, especially for high redshift clusters.
For this reason and because each mass estimator carries some key informations,
more commonly the scaling between pairs of more easily observable 
mass-related quantities is studied, such as X-ray luminosity, temperature 
or the $Y_X$ (Kravtsov et al. 2006) parameter, or optical richness.
These studies often look for outliers,
however their search is blessed by data limitation: for example 
in the search of clusters X-ray dim for their optical richness,
Donahue et al. (2001) and Gilbank et al. (2004), both mostly
worked with putative clusters (i.e. not spectroscopically 
confirmed) and X-ray undetections. 

\begin{figure*}
\psfig{figure=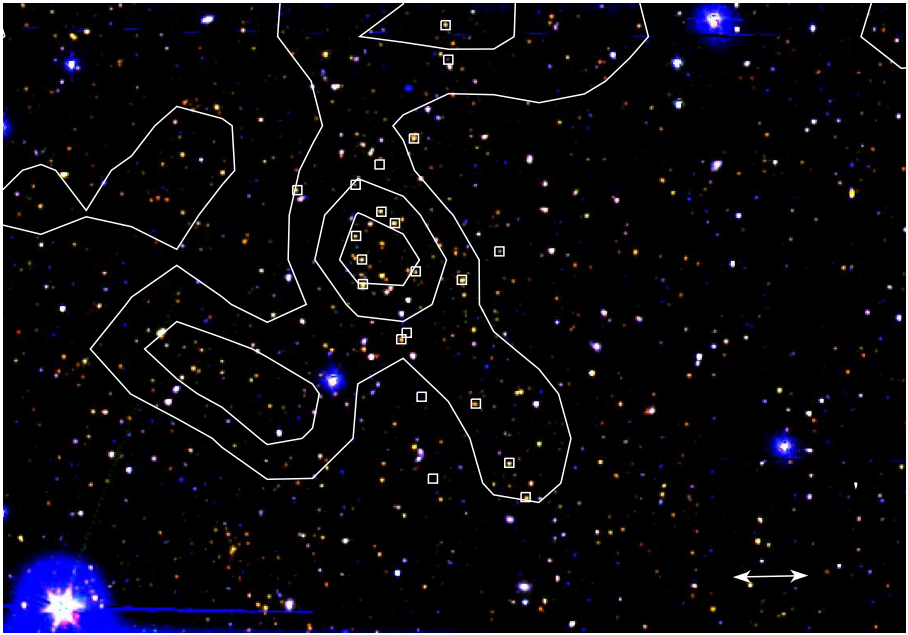,width=12truecm,clip=}
\caption[h]{True-color ($z'[3.6][4.5]$) degraded-resolution (to
make galaxies not too small when printed) image of a region of
a 24 Mpc$^2$ area around RzCS 052. Spectroscopically
confirmed clusters and isodensity contours for red
galaxies are also marked. Note the number
density contrast of reddish galaxies between the cluster center and
the right part of the image. The ruler is 1 arcmin long;
North is up, East is to the left.}
\end{figure*}

Only few works directly address the relative quality of
different mass estimators with 
velocity dispersion: Borgani \& Guzzo (2001) compare the scatter
of two mass estimators, X-ray luminosity and richness, and found that
the former is a better mass tracer than the latter when the former is
uniformly measured and the latter is taken from a 50 years old
paper reporting eye-estimate of the cluster optical richness  
(the Abell 1958 catalog). In both CNOC and nearby clusters,
mass correlates better with richness
than with X-ray luminosity (Yee \& Ellinson 2003; Popesso et al. 2005). 
Eke et al. (2004) found that optical luminosity is a better proxy
of mass than velocity dispersion in common conditions, i.e.
when velocities are available for a small sample of galaxies. 
A related issue, which we will examine below, 
is whether it exists clusters X-ray dim for
their mass (velocity dispersion), (e.g. Lubin, Mulchaey \& Postman 2004,
Fang et al. 2007, Johnson et al. 2006).

The relation between richness and mass has received some recent attention in
the form of the halo occupation function (Berlind \& Weinberg 2002; Lin et
al. 2004 and references therein) whose first moment is the halo occupation
number (HON), the average number $N$ of galaxies per cluster of mass $M$. 
In order to address the evolution of the HON,
velocity dispersion information is often unavailable for a large
cluster sample, mass and cluster size are inferred from other mass-related
quantities (for example the X-ray temperature), and assumed to evolve 
self similarly. The
evolution of the HON with redshift is still unclear: 
the initial study by Lin et al. (2004)
claimed that the HON increases at high redshift,
but Lin et al. (2006) find evidence that it does not evolve strongly out to
$z \sim 1$, suggesting that the galaxy population in clusters was established
and assembled at early epochs. Muzzin et al. (2007) confirms the above,
with a sample of reduced redshift leverage and hence reduced evolution
sensitivity, but available velocity dispersion information.

\begin{figure*}
\centerline{\psfig{figure=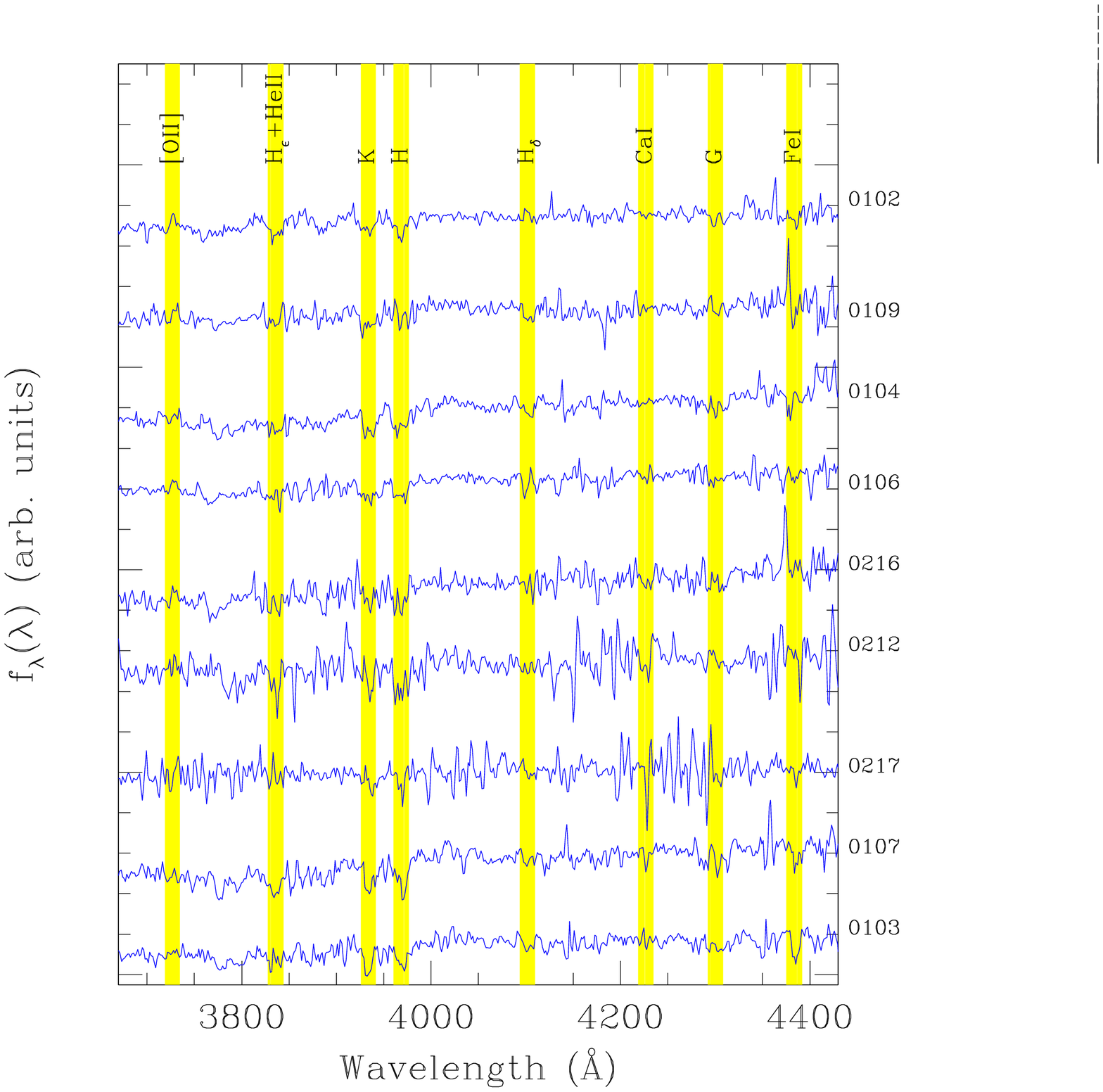,width=8truecm,clip=}	
\psfig{figure=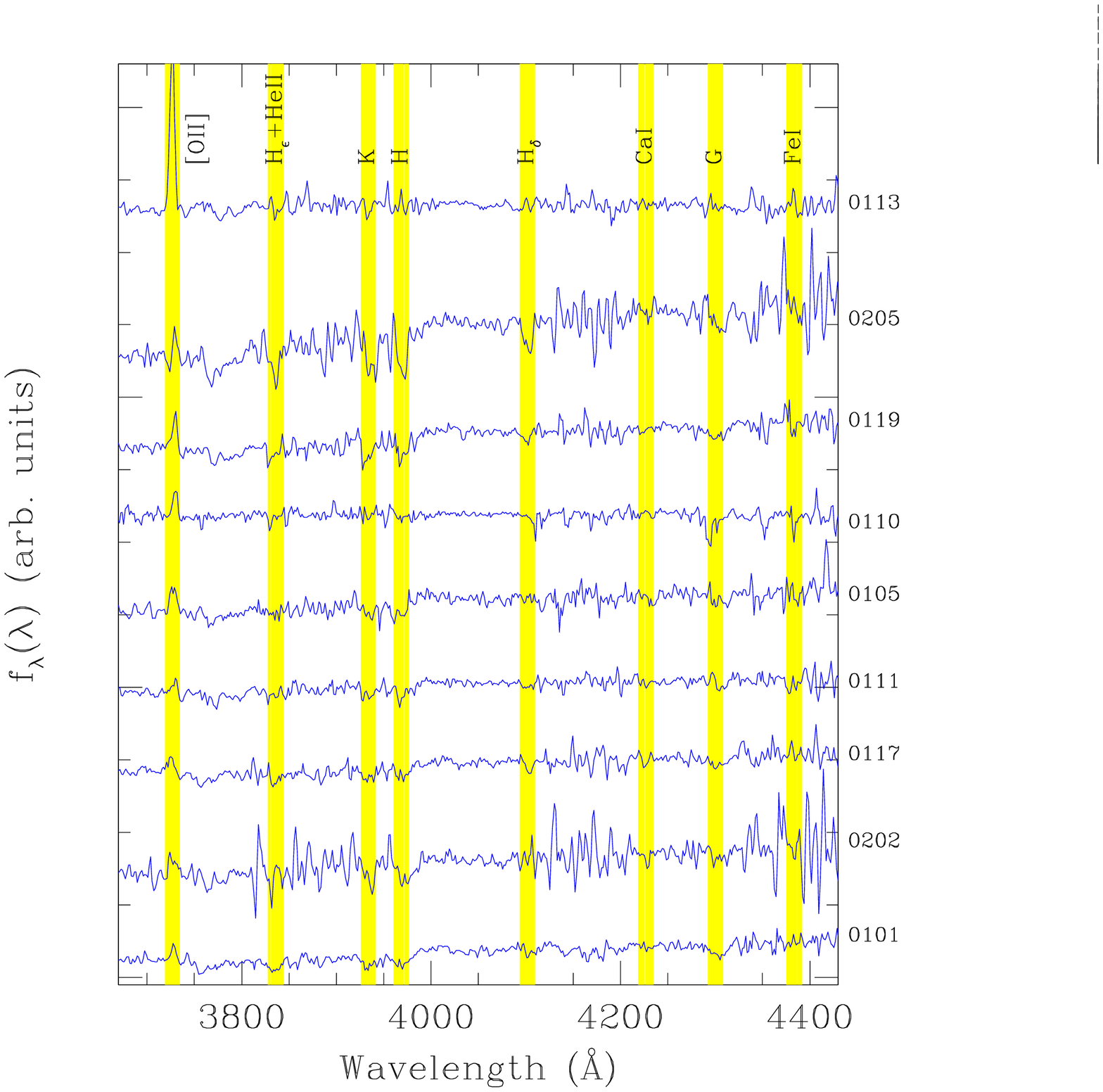,width=8truecm,clip=}}	
\caption[h]{Spectra of RzCS 052 members coming the VLT run. 
We have vertically shifted the spectra and
zoomed on a reduced wavelength range for display purpose.}
\end{figure*}

\begin{figure*}
\psfig{figure=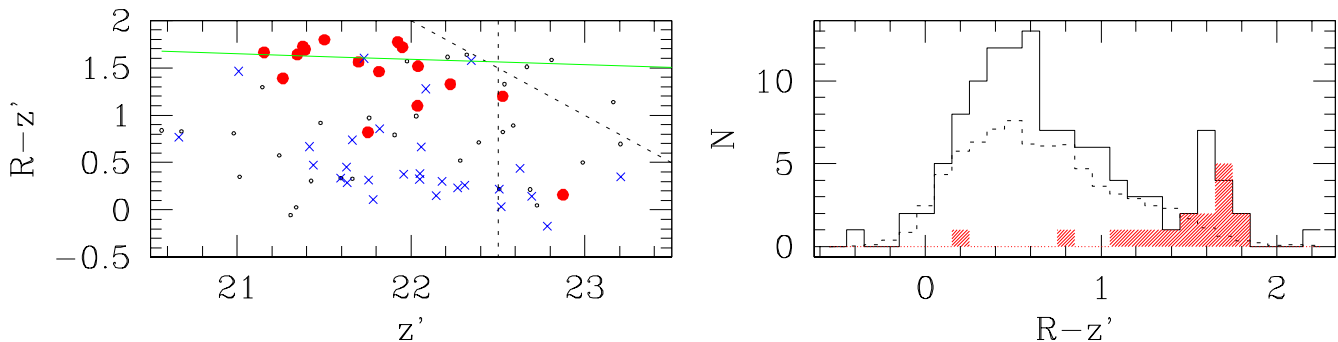,width=18truecm,clip=}
\caption[h]{
{\it Left:} Colour--magnitude diagram for galaxies (close circles) within
1 arcmin from the cluster center or with a known redshift (red closed
circles for members, blue crosses for interlopers) using CTIO
discovery data. $R$ and $z'$ mag completeness
limits are show with dashed lines.
The green line is the expected CM at the cluster redshift,
from Kodama \& Arimoto (1997) 
{\it Right:} Colour histograms of galaxies within
2 arcmin from the cluster center (solid histogram), and of the average
control field (measured on a 0.36 deg$^2$ area, dashed histogram), normalized to the
cluster area. A clear excess is seen, especially at $R-z>1.5$ mag.
The hashed histogram is the colour distribution of spectroscopically
confirmed member galaxies. In both panels, a few objects with spectroscopic
redshift are missing because their fall on bad CCD regions or have extreme
colours.}
\end{figure*}

Here, we present the photometric discovery, spectroscopic confirmation and
X-ray properties of a new $z=1.016$ cluster of galaxies (RzCS 052), a
cluster optically rich but undetected in the XMM-LSS survey (Pierre et al.
2007), and hence possibly X-ray dark (i.e. dim for its mass). We 
derive 
its global properties (richness, X-ray luminosity, velocity dispersion and
mass) and study these in the context of cluster scaling relations ($L_X -
\sigma$, HON) at high redshift. In particular, we test the claim that the
HON (the way galaxies populate cluster-scale haloes) has not changed $z \sim
1$ (Lin et al. 2006) under
far less assumptions than the original claim. We also present a Bayesian approach
to the determination
of cluster velocity dispersion and X-ray luminosity and use it to critically
examine recent claims about the existence of underluminous X-ray clusters. 
A companion paper (Andreon et al. 2007) addresses the use of RzCS 052 
as a laboratory for studying galaxy formation and evolution.

We adopt $\Omega_\Lambda=0.7$, $\Omega_m=0.3$ and $H_0=70$ km s$^{-1}$
Mpc$^{-1}$. Magnitudes are quoted in their native photometric system 
(Vega for $R$, SDSS for $z'$).

\section{The data \& data reduction}

\subsection{Photometry: CTIO $Rz'$ images}

Broadband images for a 7 deg$^2$ region around this cluster were obtained
at the Cerro Tololo Inter-American Observatory Blanco 4m telescope in the
$R$ and $z'$ ($\lambda_c\sim9000${\AA}) filters using the Mosaic II camera.
We use the same imaging data as Andreon et al. (2004a), where details
on the data and their analysis are found. Briefly, Mosaic II is a $8192 
\times 8192$ CCD camera with a $36'$ field of view at prime focus. Exposure
times were 1200s in $R$ and 1500s in $z'$: seeing was between $0.9''$ and
$1.0''$ in the final images. 

Source detection and photometry were carried out using SExtractor (Bertin 
\& Arnouts 1996). Colours and magnitudes were computed within a fixed $2''$ 
(radius) aperture, and corrected for minor differences in seeing, as in 
Andreon et al. (2004a). Completeness magnitudes ($5\sigma$ in a $3''$ aperture), 
computed as in Garilli, Maccagni \& Andreon (1999) are $R=24.0$, $z'=22.5$ mag. 

Figure 1 shows a true--colour image of RzCS 052, as derived from CTIO $z'$ 
image and IRAC Spitzer [3.6] and [4.5] images. Spitzer data reduction is
described in Andreon (2006a), that also presents 
the composite stellar mass function and the 3.6 $\mu$m luminosity 
function of many clusters, including RzCS 052.

\subsection{Spectroscopy}

Multiobject spectroscopy was carried out on Gemini in late 2003, and VLT 
in late 2003 and during 2004. On VLT, the spectra were taken using FORS2 
with the GRIS\_300I and the OG590 filter for a total integration time of 
11 ks.
On Gemini, the spectra were obtained with the Gemini Multi Object Spectrograph
(GMOS), operating in nod \& shuffle mode (Abraham et al 2004; Cuillandre et al.
1994) in order to perform accurate sky subtraction, with the R150
grating for a total integration time of 15 ks.

The Gemini GMOS package for IRAF was used to calculate the wavelength
solutions and to reduce the multiobject observations into one-dimensional
spectra. The RVSAO package (Kurtz \& Mink 1998) was used to measure redshifts
(and their errors) of target galaxies by cross-correlation with stellar and 
galaxy templates of known radial velocity (Tonry \& Davis 1979). 

A total of 57 spectra of 54 galaxies yielded reliable redshifts,
with typical individual errors on redshift of 50 to 150 km/s
(depending on instrument, exposure time, object spectrum, etc.). 
Three galaxies with duplicate observations have concordant redshifts
in the two data sets. Figure 2 shows the spectra of RzCS 052 members
from the VLT run. Table 1 list position and redshift of galaxies
within 4000 km s$^{-1}$ of RzCS 052.

\begin{table}
\caption{J2000 coordinates and redshift of
galaxies within $4000$ km s$^{-1}$ (rest-frame)
of RzCS 052}
\begin{tabular}{ccc}
\hline
 RA  &  DEC  & redshift  \\
\hline
\hline                                                   
02:21:36.21 & -03:24:56.0 & 1.0210 \\
02:21:37.08 & -03:24:28.4 & 1.0192 \\
02:21:37.60 & -03:21:38.0 & 1.0176 \\
02:21:38.85 & -03:23:40.7 & 1.0206 \\
02:21:39.60 & -03:22:00.9 & 1.0217 \\
02:21:40.32 & -03:19:03.4 & 1.0225 \\
02:21:40.46 & -03:18:35.6 & 1.0158 \\
02:21:41.13 & -03:24:41.2 & 1.0195 \\
02:21:41.73 & -03:23:35.2 & 1.0089 \\
02:21:42.04 & -03:21:54.1 & 1.0132 \\
02:21:42.14 & -03:20:07.0 & 1.0074 \\
02:21:42.52 & -03:22:43.6 & 1.0156 \\
02:21:42.81 & -03:22:48.8 & 1.0181 \\
02:21:43.15 & -03:21:15.2 & 1.0065 \\
02:21:43.87 & -03:21:06.0 & 1.0129 \\
02:21:43.96 & -03:20:27.9 & 1.0159 \\
02:21:44.85 & -03:22:04.3 & 1.0230 \\
02:21:44.90 & -03:21:44.5 & 1.0145 \\
02:21:45.21 & -03:21:25.5 & 1.0187 \\
02:21:45.24 & -03:20:44.3 & 1.0151 \\
02:21:48.33 & -03:20:48.6 & 1.0160 \\
\hline 
\end{tabular}                               
\end{table}

\begin{figure}
\psfig{figure=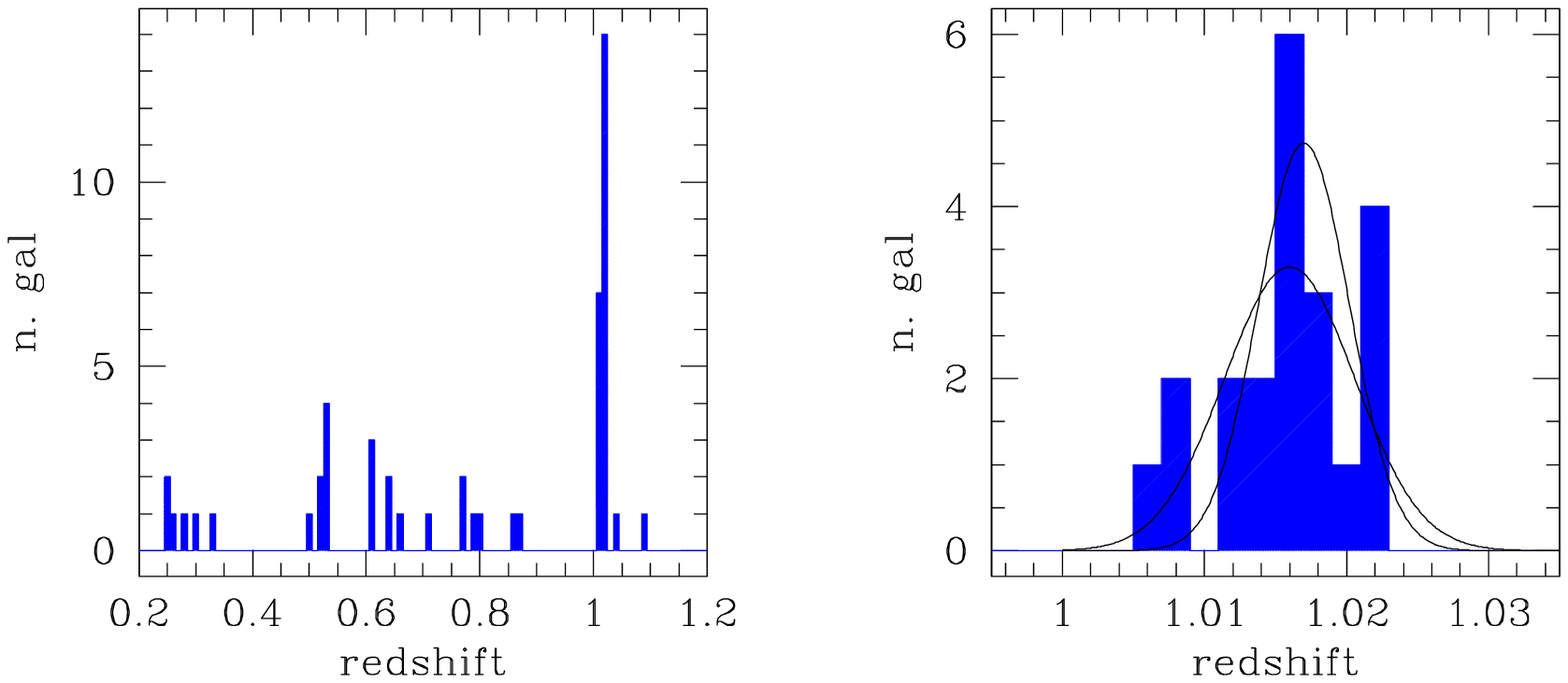,width=8truecm,clip=}
\psfig{figure=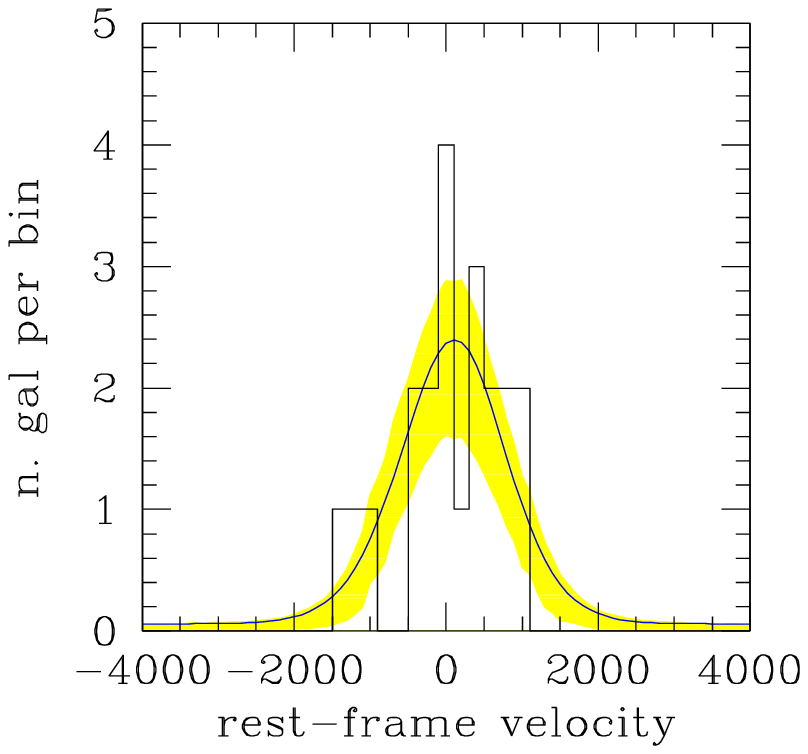,width=8truecm,clip=}
\caption[h]{Redshift distribution of all successfully measured
redshift in the cluster line of sight (top panel) and around
the $z=1.016$ (bottom panel). In the bottom panel, the curve mark
the mean model, and the gray (yellow, in colour) region is the
68 \% highest density posterior interval.}
\end{figure}

\subsection{XMM-Epic data}

RzCS 052 was observed with {\it XMM Newton} using the European Photon Imaging 
Camera (EPIC) instrument (Jansen et al. 2001) in 2002 in full-frame mode with
the thin filter. After flares filtering, the good exposure time is $\sim$13 ks 
for the MOSes  (Turner et al. 2001), and a $\sim$8 ks for the PN (Struder et al. 
2001). By using the {\it XMM-Newton} Science Analysis System (SAS, v. 7) package 
and our own scripts, we kept only patterns between 0 and 12 for MOS and 0 to 4 for 
PN. We flagged bad pixels, bad columns and CCD gaps, regions not seen by all three
instruments, as well pixels contaminated by the flux of other sources. We remove the 
energy band [0.60-0.70] keV, where an instrumental line shows up because this flattens 
the sky background, and hence decreases the complexity of the model
used to describe its spatial distribution. We merged the three instruments 
to improve S/N. 

For comparison, we also reduced EPIC observations of a cluster at almost identical 
redshift, XLSSC 029 at $z=1.05$ (Andreon et al. 2005), just one degree apart from 
RzCS 052. To make the comparison straightforward we cut the XLSSC 029 exposure to 
match (almost) exactly the exposure time of  RzCS 052. 

\section{Results}

\begin{figure}
\psfig{figure=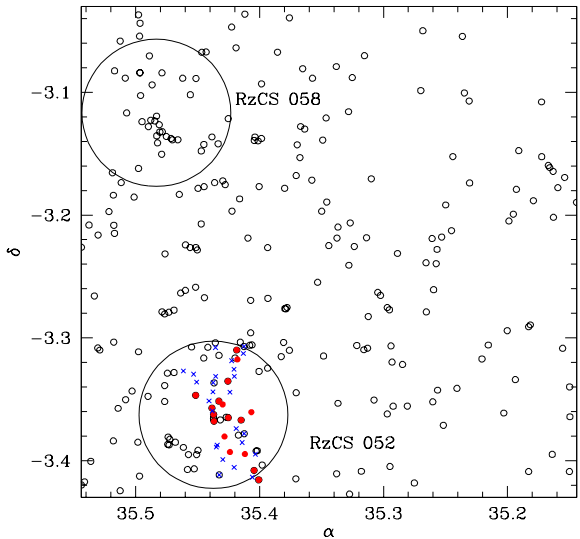,width=8truecm,clip=}
\caption[h]{Spatial distribution of red galaxies (open points),
spectroscopically confirmed members (solid red points), and spectroscopically 
confirmed inderlopers (blue crosses).}
\end{figure}

\subsection{Photometric discovery and colour--magnitudes relation} 

RzCS 052 was been initially detected in 2000 using  photometric data ($Rz'$) as
a clustering of sources of similar colour using our own version  (Andreon 2003;
Andreon et al. 2004a,b) of the red-sequence method (Gladders \& Yee 2000). This
is shown in the right-hand side panel of Figure 2, which shows that in a $2'$
circle centered on RzCS 052 (02:21:42 $-$03:21:47 J2000) there are $13$ galaxies
redder than $R-z'=1.55$ mag (solid histogram), while  the expected number in the
same area (i.e. background, average number measured in a 0.36 deg$^2$ area all
around the cluster) is $\sim2.1$, a very obvious overdensity detection.

The colour of the red sequence (Figure 2), compared to those of other high 
redshift clusters presented in Andreon et al. (2004a) suggests a redshift of
$z\sim1.0$. F. Barrientos (2006, private communication) confirmed that this
cluster has also been detected by their red-sequence cluster survey 
with which we share the CTIO imaging.

The right panel of Fig 2 shows that the colour distribution of
RzCS 052 galaxies (the area between the solid and dashed histogram)
is bimodal, displaying a narrow peak at $R-z'=1.6$ mag and a broad
excess at bluer colours. From now on, we define galaxies as red if
$1.4<R-z'<1.9$ mag. Isodensity contours for red galaxies are shown in
Fig.~1

\subsection{Spectroscopical confirmation and velocity dispersion}

The upper panel of Figure 4 shows the distribution of successfully measured
redshifts in the cluster line of sight. The clear peak at $z\sim1.02$ 
is in good agreement with the photometric redshift inferred from the colour of
the red sequence ($z\sim1.0$). The lower panel shows a detailed view around 
the cluster redshift. We
measure $z_{cluster}=1.016$ and $\sigma_v=710 \pm 150$ km s$^{-1}$ (see
Appendix B for methods). The gapper or biweight estimators (Beers et al. 1990)
give identical $\sigma_v$.

\subsection{Richness}

Figure 5 shows the spatial distribution of red galaxies (open points)
in a wide area of 133 Mpc$^2$ around RzCS 052. Two galaxy overdensities
are quite obvious, both colour-detected by our cluster detection algorithm. 
Spectroscopically confirmed RzCS 052 members (solid red points) and 
spectroscopically confirmed interlopers (crosses) are also marked. 
The two large circles have a radius of $3.6'$, which is 80 $\%$ of
the Abell (1958) radius (at the RzCS 052 redshift).

We fit a $\beta$ profile to the distribution of galaxies (Appendix A) and
remove contamination using counts from the field in Fig.~5, discarding the
region around RzCS 058. Within one Abell radius we find $56 \pm 20$ red cluster
galaxies brighter than $z'=22.5$ mag. This number must be corrected to
$M_3+2$ using the luminosity function and the 30$\%$ blue fraction measured
in Andreon et al. (2007). The total number of galaxies is $\sim 150$, that 
qualifies RzCS 052 as an Abell richness class 3. A different richness estimate
is presented in sec 4.

\subsection{X-ray luminosity}

The RzCS 052 cluster is within the XMM Large Scale Structure
Survey (LSS) field, but not X-ray detected by the current XMM-LSS pipeline
(Pierre et al. 2007), even thought several other
$z\sim 1$ clusters are (Valtchanov et al. 2004, Andreon et al. 2005,
Bremer et al. 2006 and some more yet unpublished). 

The left panel of Fig 6 shows the X-ray image of RzCS 052. The X-ray
source close to the optical cluster centre is not extended 
(Figure 7) and appears
to coincide with a foreground spiral (as classified from Hubble
Space Telescope images presented in Andreon et al. 2007) galaxy, and
thus not associated with RzCS 052. Therefore, the flux from this source 
is discarded in the determination of the X-ray flux of RzCS 052.

\begin{figure*}
\psfig{figure=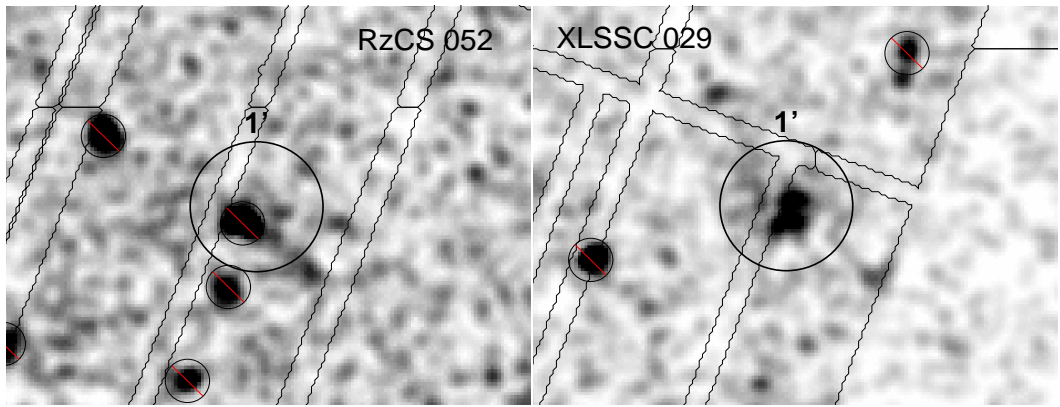,width=14truecm,clip=}
\caption[h]{[0.5-2.0] keV image of RzCS 052 (left panel) and XLSSC 029
(right panel), at very similar redshift and with matched X-ray images
and smoothing. Pixels affected by other sources, or falling on CCD gaps
are marked by regions. Simple eye inspection confirms that RzCS 052 is
much fainter than XLSS 029.}
\end{figure*}

\begin{figure}
\centerline{\psfig{figure=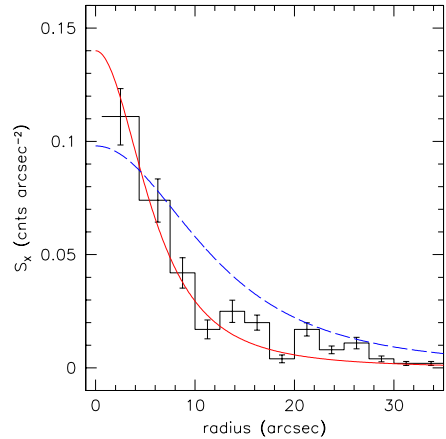,width=6truecm,clip=}}
\caption[h]{Radial profile of the foreground X-ray point source (black solid 
histogram) and of the XMM PN PSF (red continuous line), at
the off-axis angle of the source, and a $\beta=2/3$ model having
$r_c=16$ arcsec (150 kpc if at the cluster distance, dashed blue line). 
The source is unresolved at the
XMM resolution, the slight excess with respect to the PSF being
due to the unremoved contribution of RzCS052, and far more compact
than a $\beta$ model with a typical cluster core radius at $z=1$ . }
\end{figure}

The right-hand panel of Fig.~ 6 shows an X-ray image of a very similar 
(in redshift, off-axis angle and exposure time of observations) cluster: 
XLSSC 029.  XLSSC 029 is much brighter than RzCS 052. 

In order to determine the X-ray luminosities of XLSSC 029 and RzCS052, we use 
the method described in Appendix A. We assume uniform priors, zeroed in
the unphysical ranges (negative core radius, negative background intensity,
negative central cluster intensity) and in ranges that make the total
flux infinity (i.e. for $\beta<0.4$). Besides returning a flux uncertainty that
account for the covariance of all parameters,  we also account for the  
cluster flux in the background region 
(for XLSSC 029 this turns out to lead to an underestimate of its count rate by
30$\%$). 

For XLSSC 029 we found: $L_X [1-4]$ keV band: $4.4\pm0.8 \ 10^{44}$ erg s$^{-1}$, 
formally for a temperature of $4$ keV (taken from Pierre et al. 2006), but actually 
for a  range of temperatures because of our choice of quoting luminosities in  the
[1-4] keV band, i.e. in a band that, at the cluster redshift, matches the observer 
frame [0.5-2] band, and because of the very tiny dependency of the conversion factor on
temperature.  For RzCS 052 we found: $f_X=1.2\pm0.8 \ 10^{-12}$ erg s$^{-1}$ cm$^{-2}$ 
and [1-4] keV rest-frame band $L_X=0.68\pm0.47\  10^{44}$ erg s$^{-1}$  
(both values are posterior mean and standard deviation). The cluster is not
an $\approx 1.5$ detection, however: the posterior probability $p(f_X<f_0|data)$
goes to zero at fluxes  $f_0\ga 1 \ 10^{-13}$ erg s$^{-1}$ cm$^{-2}$, 
i.e. the source has so
many detected photons that data cannot be described by a model with a
cluster signal fainter than
$f_0$, such as a model including background emissivity only. 
The large flux and intensity uncertainties account 
for the uncertainty
of the beta function parameters (core radius, beta and central intensity)
and background value.
 
\subsection{$L_X-\sigma_v$ relation}

Figure 8 shows the location in the $L-\sigma_v$ plane of
RzCS 052 cluster, with all the $z>0.8$ clusters for which we found
in literature $T, L_X$ and $\sigma_v$ (Table 2). We ignored a few 
tentative $\sigma_v$ 
determinations based on small number of velocities, because affected by
large errors and by
Eddington bias (detailed in appendix B and in sec 3.6). 
Literature values of $L_X$ 
are converted to the [1-4] keV band rest-frame to minimize systematics.

\begin{figure}
\psfig{figure=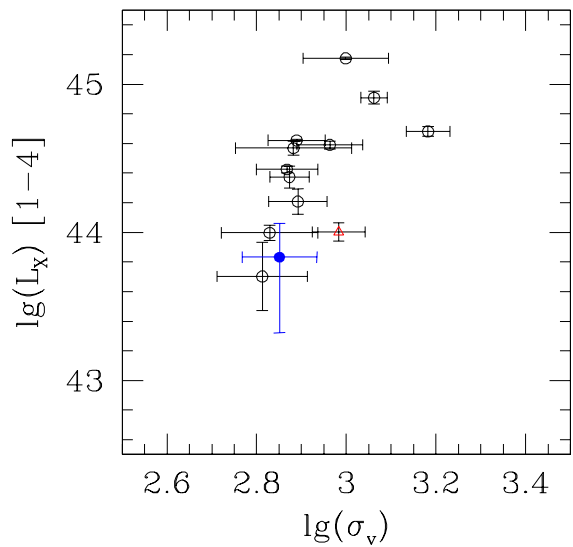,width=8truecm,clip=}
\caption[h]{$L_X-\sigma_v$ relation for literature clusters (open points)
and for RzCS 052 (close point). The (red) triangle is CL1604+4304, i.e. the cluster
originally claimed to be X-ray dark, at the revised value of the $\sigma_v$ 
determination (see text for details).}
\end{figure}

\begin{figure}
\psfig{figure=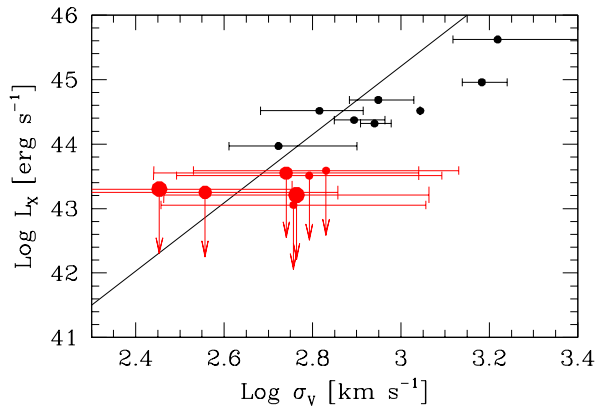,width=8truecm,clip=}
\caption[h]{$L_X-\sigma_v$ relation for literature clusters (close black 
points), as published by Fang et al. (2007),
and DEEP2 groups and clusters, after our revision (red points and arrows).
DEEP2 groups and clusters are not longer underluminous for their
velocity dispersion.}
\end{figure}

RzCS 052 (closed circle) turns out to be one of the faintest
and least massive (lower velocity dispersion) clusters known at $z>0.8$.
In spite of the cluster being optically selected, RzCS 052 has
an X-ray luminosity appropriate for its mass (velocity dispersion),
or at least, data are compatible with the trend seen for similar,
but X-ray selected, clusters. Therefore, although the X-ray luminosity of 
RzCS 052 is modest, it is consistent with measurements for other high 
redshift clusters. This is reassuring for ongoing X-ray cluster surveys, 
which usually assume a Gaussian model for the scatter between mass and X-ray 
luminosity: RzCS 052 is not an example of a new class of
clusters, massive but dark in X-ray, easily escaping 
the detection in X-ray surveys because faint for
their mass. Instead, RzCS 052 has an X-ray luminosity appropriate 
for its mass (velocity dispersion) and it is
missed in the XMM Large Scale Structure 
because the low survey sensity at $z\sim1$ for
objects of RzCS 052 $\sigma_v$ and obeying to the $L_X - \sigma_v$ relation.

\subsection{Are there known optically underluminous clusters or groups at
high redshift?}

With the Bayesian tools described in the Appendix, 
we revisit claims about the existence of underluminous
X-ray clusters (i.e., clusters whose masses -- from their velocity dispersions
-- are too large for their X-ray luminosity -- which is often an upper limit).
We have seen (Appendix B) that in the case of Cl1604+4304 a Bayesian estimate of its
velocity dispersion, as well as the revised value of $\widehat{\sigma_v}$ published by
Gal \& Lubin (2004), make this cluster no longer an outlier (i.e., underluminous)
in the $L_X - \sigma_v$ relation. We show here that Bayesian estimates of
velocity dispersions and X-ray fluxes cast serious doubts on the existence
of the X-ray underluminous groups or clusters claimed in literature.

Fang et al. (2007) study seven DEEP2 groups at $0.75<z<1.03$. They derive
an upper limit to the X-ray flux by considering only photons within an
aperture of radius $30''$, a metric radius of 250 kpc which is considered
to be typical for groups and clusters at high redshift. However, assuming
$\beta=2/3$ (also a typical value), the X-ray flux outside of their aperture,
integrated to infinity, is 2.4 times larger then the flux inside their
aperture. The upper limit quoted by Fang et al. (2007) is therefore too
small by a factor of 3.4. 

Fang et al. (2007) derive their velocity dispersion from 3 to 6
galaxies. Because of Eddington (1940) biases, this is biased high: even
symmetric errors move more low velocity systems to high velocity dispersion
than otherwise. The Bayes theorem allows to correct for the bias, being
the Eddington correction built in the Bayes theorem (Appendix B).

If these sources of error and biases are accounted for, the X-ray fluxes and velocity
dispersions for the groups studied by Fang et al. (2007) are perfectly
consistent with the local $L_X - \sigma_v$ relation (Figure 9). We note here that a
similar argument can be made for the `underluminous' CNOC groups claimed
by Spiegel, Paerels, \& Scharf (2007), based on velocity dispersions
computed on just 3 or 4 velocities, and upper limits on the X-ray luminosity.

Popesso et al. (2007) also claim that there exist X-ray
underluminous clusters in the local universe, but their definition of `underluminous'
depends on the data depth. Most of their
underluminous clusters have normal X-ray
luminosity for their mass, because they obey to
the $L_X - \sigma_v$ relation (see their figure 2d),
and are called `underluminous` because
their are faint in their X-ray imaging. Deeper data would have
classified them as `normal'. The few
remaining clusters are found to have a negative (unphysical) X-ray flux
and seem underluminous in
the $L_X-\sigma_v$ relation because they are plotted at an arbitrary
value of $L_X$ rather than as an upper limit. Such objects are not 
underluminous in the generally understood sense.


To summarize, to our best knowledge there is no evidence for not even
a single example of cluster (or group) of galaxies X-ray dim for its
velocity dispersion (once all sources of errors are accounted for),
all previous claims proven to be based on uncertain grounds.
If there are underluminous clusters, they have not yet been 
convincingly discovered.

\begin{figure*}
\psfig{figure=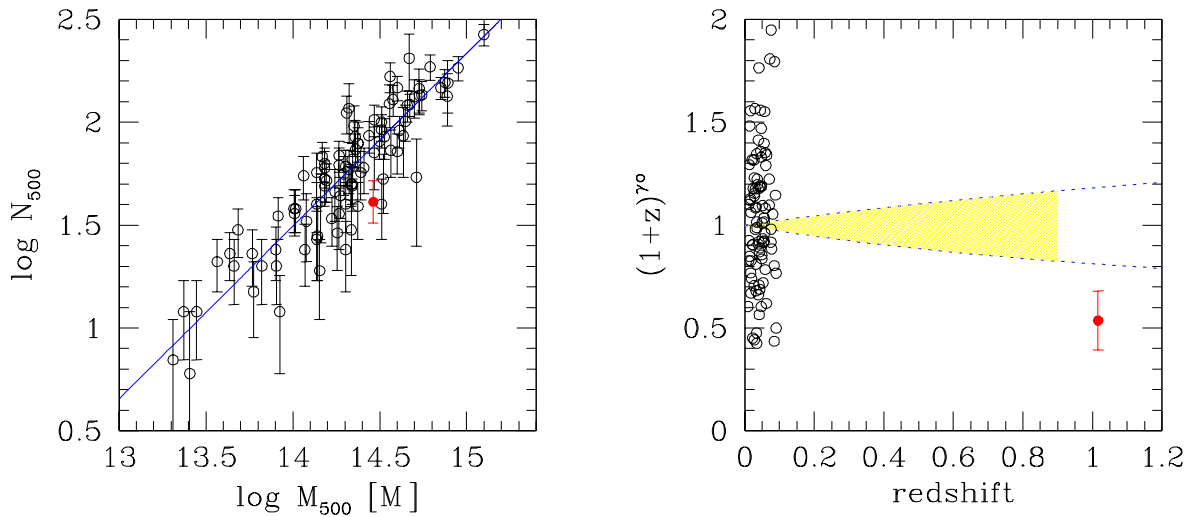,width=16truecm,clip=}
\caption[h]{{\it Left panel:} Number of galaxies as a function of cluster mass.
Open points mark local ($z<0.1$) clusters, the closed (red) point mark RzCS 052.
The line is the Lin et al. (2006) fit to local clusters.
{\it Right panel:} Evolution of the $N-M$ relation. Points are as left panel. The
shaded (yellow) region mark the 68 \% confidence region derived by Lin et al. (2006)
at $z<0.9$.}
\end{figure*}

\section{Halo Occupation Number}

We now derive the HON for RzCS 052. This basically requires an estimate
of the $N$ galaxies within a specified luminosity range and within a 
given cluster-centric radius and a mass estimate within the same 
spherical volume.

The virial radius, $r_{200}$, and mass, $M_{200}$ are derived from the 
virial theorem from the measured $\sigma_v$: $r_{200}=1.04$ Mpc, and
$M_{200}=4.0\ 10^{14} M_\odot$. Adopting a Navarro, Frenk \& White (1997)
profile with concentration $c_{dm}=5$ we derived $r_{500}=0.69$ Mpc 
and cluster mass within $r_{500}$: $M_{500}=2.9\ 10^{14} M_\odot$.

The number of galaxies is computed by integrating the luminosity function 
(LF, hereafter) within $r_{500}$ down to $M^*+3$. The latter is derived from 
the deeper VLT $z'$ data presented in Andreon et al. (2007). 
Since the latter work count 
red galaxies only in their LF, we correct for the blue galaxies
fraction (adopting the blue fraction measured in Andreon et al. 2007). We
also correct for the distribution of cluster members
in a `cylinder' outside the cluster sphere, assuming a
NFW distribution with concentration $c_{gal}=3$ (as in Lin et al. 2004, 
2006). We opted for a Bayesian approach, because it 
simplify the computation of the uncertainty on the cluster richness 
fully accounting for uncertainties and covariances (neglected in
past works) for all (Schechter and background) parameters.
We found: $N_{500}=41 \pm11 $ galaxies. Fig.~10 compares RzCS 052
to the local richness-mass scaling, showing that it is within the local
relation, although near the bottom-end of the distribution.

Following Lin et al. (2006) we parametrize the evolution of $N(m,z)$ 
as:

\begin{equation}
N(M,z)= N_0\ (1+z)^\gamma \ (M/M_0)^s  
\end{equation}

where $N_0=56$ and $M_0=2 \ 10^{14}$ are normalization factors of the relation, 
and $s=0.84$ is the slope of the local relation derived by Lin et al. (2004)
for their local cluster sample. We can rewrite this as below, to emphasize the
evolutionary terms:

\begin{equation}
(1+z)^{\gamma_0} = \frac{N_0(M,z)}{N_0 \ (M/M_0)^s}     
\end{equation}

where we have added the subscript $0$ to emphasize that we are now talking
about the observed (or maximum likelihood) values. In order to estimate
$\gamma$ we can just look at the dependence of the r.h.s of eq 2 with $(1+z)$, as 
shown in the right-hand panel of Fig.~10. The value observed for RzCS 052 
is a bit farther away in redshift than the range probed by Lin et al. (2006), and
is outside their 68 \% interval on $\gamma$, shown as shaded (yellow)
area. A zero value would implies that the way galaxies populate cluster-scale halos 
at $z=0$ has not changed from $z=1$. The size of the error bar on RzCS 052 is
comparable to the error on $\gamma$ (the width of shaded region in figure at $z=1.0$), indicating
that RzCS 052 alone carries comparable information to all the high redshift
clusters studied by Lin et al. (2006). Therefore, the data for RzCS 052 suggest a 
mild evolution, with the caveat that the scatter around the mean relation is large 
(see left-hand panel) and our result should be taken as tentative. We note,
however, that our measurement is more direct than Lin et al. (2006): we
include an estimate of the characteristic luminosity $M^*$ and faint-end slope 
$\alpha$ from our data, whereas Lin et al. (2006) assumed them for lack of data,
and we measure mass and reference radius, $r_{500}$, from the virial theorem
without assuming that they scale with X-ray temperature and evolve self-similarly, 
as assumed by Lin et al. (2006) for lack of direct measurements.
The mild difference seen in the right panel of Fig 10 may indicate a possible
break in the (assumed) self-similar evolution of the scaling between temperature and 
radius or mass at $z\sim 1$.  
 
The parameter $\gamma$ is the (logarithm) derivative of the redshift dependence 
of the number of galaxies per unit cluster mass, i.e. of the 
galaxy merging rate in appropriate
units. Lin et al. (2006) and our results agree that the number of galaxies 
per unit cluster mass has increased (us) or stayed constant (Lin et al. 2006) 
since $z=1$. Therefore, both studies directly show that no 
intense merging activity of galaxies has been ongoing in clusters in the last 
7 Gyr.

\section{Summary}

We have identified a distant cluster from a modified red-sequence method
and followed it up spectroscopically. RzCS 052 is a richness class 3
cluster at $z=1.016$ with a velocity dispersion of $710 \pm 150$ km s$^{-1}$
and an X-ray luminosity of $0.68 \pm 0.47 \times 10^{44}$ ergs s$^{-1}$. 

In spite of its optical detection,
RzCS 052 obeys to the high redshift $L_X-\sigma_v$ relationship as other 
X-ray selected clusters
to the high redshift $L_X-\sigma_v$ relationship, 
whereas in principle variations in the dynamical state 
of the clusters
or in the thermal history of the intracluster medium may have moved
it away from the $L_X-\sigma_v$ relation.

Analysis of the $N-M$ scaling shows that RzCS 052 
has the right number of galaxies (actually, a bit less) that it should have for its mass, 
ruling out intense merging (among galaxies) activities
in clusters from $z=1$ to today, in agreement with Lin et al. (2006).

We present a Bayesian
approach to measuring cluster velocity dispersions (most useful for sparsely
sampled data and in presence of a background)
and X-ray luminosities or upper limits (essential 
in the case of poorly determined parameters).
Critical re-analysis of the data of 
clusters/groups claimed to be outliers of the $L_X-\sigma_v$ relationship leads
to conclude that there are no known thus far examples of clusters X-ray underluminous 
for their velocity dispersion. The above result is quite reassuring for
ongoing X-ray surveys: there is thus far no example of cluster missed because
an anomalous $L_X$ for the cluster mass.

\begin{table*}
\caption{Luminosity and velocity dispersion of clusters at $z>0.8$}
\begin{tabular}{llcrcrr}
\hline
Name & z & $\log L_X [1-4]$ & ref & $ \sigma_v$ & N \hfill & ref \\
\hline
RXJ1716+6708   & 0.813 & $44.68\pm0.03$ & 12 & $1522 \pm 180 $& 37 & 1 \\
RXJ1821.6+6827 & 0.816 & $44.62\pm0.01$ & 2 & $775  \pm 122 $& 18 & 2 \\
MS1054-30321   & 0.830 & $44.91\pm0.04$ & 12 & $1153 \pm 80  $& .. & 3 \\
RXJ0152-1357S  & 0.830 & $44.42\pm0.02$ & 12 & $737  \pm 126 $& 18 & 4 \\
RXJ0152-1357N  & 0.835 & $44.59\pm0.03$ & 12 & $919  \pm 168 $& 16 & 4 \\
RzCS 530       & 0.839 & $44.20\pm0.09$ & 5 & $780  \pm 126 $& 17 & 5 \\
1WGA1226+3333  & 0.890 & $45.17\pm0.01$ & 12 & $997  \pm 245 $& 12 & 6 \\
Cl1604+4304    & 0.900 & $44.00\pm0.06$ & 14,7 & $962  \pm 141 $& 67 & 7 \\
RzCS 052       & 1.016 & $43.83\pm0.37$& this work & $710 \pm 150 $& 21 & this work \\
RXJ0910+5422   & 1.106 & $44.00\pm0.05$ & 12 & $675  \pm 190 $& 25 & 8 \\
RXJ1252-2927   & 1.237 & $44.37\pm0.07$ & 12 & $747  \pm 79  $& 38 & 9 \\
LynxW          & 1.270 & $43.70\pm0.23$ & 12 & $650  \pm 170 $& 9  & 10 \\
1WGAJ2235.3    & 1.393 & $44.57\pm0.05$ & 11 & $762  \pm 265 $& 12 & 11 \\ 		    
\hline                                                   
\end{tabular} \hfill \break                              
\footnotesize{RzCS 530 is also known 
as XLSSC 003; References: 1: Gioia et al. (1999); 2: Gioia et al. (2004a);
3: Gioia et al. (2004b); 4: Demarco et al. (2005); 5: Valtchanov et al. (2004);
6: Maughan et al. (2004); 7: Gal \& Lubin (2004); 8: Mei et al. (2006); 
9: Demarco et al. (2007); 10: Stanford et al. (2001); 11: Mullis et al. 
(2005); 12: Ettori et al. (2004); 14: Lubin et al. (2004). \hfill \break
}
\end{table*}

\section*{Acknowledgments}

We thank the referee for useful comments that improve the
paper presentation.
SA thanks A. Diaferio, Ben Maughan, A. Muzzin for useful discussions. 
This paper is based on observations obtained, with CTIO
(prog. 2000-0295), ESO (072.C-0689, 
072.A-0249, 073.A-0564, 074.A-0360), Gemini (GS-2003B-Q-18)
XMM (OBS. ID. 0037980601).
We acknowledge partial financial support from contract ASI-INAF I/023/05/0
and Centro de Astrofisica FONDAP.

\bsp

\appendix

\section{Mixture modelling of inhomogeneous processes for the $L_X$ and 
richness estimates}

We want to measure a structured (i.e. not constant in space) Poisson signal 
in presence  of a background. We assume that the background (photons, galaxies, 
etc.) distribution is an homogeneous (i.e. the intensity is independent of position) 
Poissonian random process, whereas the cluster contribution is an inhomogenous 
Poissonian random process whose intensity is given by an $I(r)$ radial profile. 
Provided that quantities are Poisson distributed, it does not matter if we
are talking about X-ray photons (as in sec 3.3), or galaxies (as in sec 3.4) or 
something else.

Let us call $\theta$ the (unknown) set of parameters of the function $I(r)$. Simple 
algebra shows that the likelihood function $\mathcal{L}(\theta) \equiv p({r_i}|I(r))$ is 

\begin{equation}
\mathcal{L}(\theta) = \prod_{i} \omega(r_i) I(r_i) \ \  e^{-\int_{\Omega} \omega(r) I(r)}
\end{equation}

where $\Omega$ is the solid angle. The expression can be simplified somewhat by noting 
that the infinitesimal solid angle at $r_i$, $\omega(r_i)$, is independent of the parameter 
$\theta$ and therefore can be dropped. There are no limitation on the complexity of the 
shape of the solid angle. $\Omega$ also encodes the relative efficiency of the different 
parts of the instruments (for example, the different efficiency of off- and on- axis 
response).

Combined with prior probability distributions for the parameters, this likelihood
function yields, via the Bayes theorem, the posterior distribution for the 
function parameters $\theta$, given the data. Marchov-Chains Monte Carlo (Metropolis 1953) 
with a Metropolis (1953) sampler is used to sample the posterior. The chain provides a 
sampling of the posterior that directly gives credible intervals for whatever quantity, 
either for the parameters $\theta$ or any derived quantity such as total richness (or flux): 
for an interval at the desired credible level it is simply matter of taking the interval 
that includes the relevant percentage of the samplings. Credible intervals (yellow areas
in Fig 4 and B1) are computed in that way. 
Upper limits may be determined in the same way 
as fluxes for detection, i.e. by specifying the credible interval we are interested in.

The function $I(r)$ can be whatever function positively defined and
having a finite integral. In this paper we use a modified
$\beta$ function:

\begin{equation}
I(r) \propto [(1+r/r_c)^2]^{-3\beta+1/2} +bkg
\end{equation}

were we have accounted for a constant background, $bkg$. By 
choosing a more complex background function, as in
Andreon (2006b), we obtain the aimed mixture 
modeling of two inhomogenous Poissonian processes.

\subsection{Fallacies of the usual measurements of $L_X$ upper limits}

While our way of determining fluxes, richness and their errors, as well as
$L_X$ upper limits  is unusual in our astronomical context (but the standard 
approach in other fields of astronomy and in statistics), we were obliged to introduced it because previous
approaches are unsatisfactory when an important parameter has a large
error or is undetermined.

A common assumption of many determinations of upper limits to the X-ray flux
from a cluster is that the object flux is fully inside a given aperture or 
the object core radius and $\beta$ are known. However, this is a dangerous
assumption: if the object is undetected, its extent, core radius and $\beta$
are not constrained. If, for example, the object is much larger than assumed 
or $\beta$ is small (and data tell nothing about that, being the object
undetected) the assumption has important consequences. For example, it is
sufficient to assume that the `underluminous' groups of Fang et al. (2007) 
have `typical' core radii to make their $L_X$ compatible with the local $L_X - 
\sigma_v$ relation (and the groups no longer underluminous).

A first step in the right direction is to correct  
for the flux outside the aperture, but this assume to known the unknown:
when $r_c$ and $\beta$ are unmeasured, or are very poorly determined, 
we cannot assume them as perfectly known and we cannot make inferences
dealing with quantities strongly depending on the poorly determined parameters, 
such the location of clusters in the $L_X-\sigma_v$ relation.
The scientific method does not suggest to {\it hope} to have taken, 
by good chance, the correct value of an unknown
parameter (as $r_c$) when it strongly affects the result.

However, this is exactly the kind of problem where Bayesian approaches are
most valuable. Bayes' theorem allows us to infer the value of a quantity (in
this case $L_X$) in the presence of a nuisance parameter (core radius or
$\beta$) whose value is unknown but whose value affects the measurement of
the quantity. Assuming a single value for nuisance parameters artificially
collapses the error ellipse along one (or more) axes and leads to an incorrectly
small error bar and to call outlier something that instead is fully compatible
with the model. The sum rule of probability prescribes to marginalize
(average over) nuisance parameters, not to keep them fixed.

Other authors determine upper limit to the X-ray flux mistakenly taking 
the maximum likelihood estimate of sampling theories, $total-background$, 
for the true value of the net flux. While $total-background$ is allowed to 
be negative, the true value of the net flux cannot be. These two quantities 
differ when the net flux is comparable to background fluctuations (e.g. appendix
B of Andreon et al. 2006). 

Another common way to compute the upper limit of the X-ray flux is by measuring 
the fluctuations of background counts. While this number is interesting in its own
right and has the appealing property that it becomes smaller and smaller with lower 
and lower background fluctuations, it is measuring something different than
the X-ray flux. In fact this quantity is a $p-$value, i.e. a measure of how
frequently one observes larger background fluctuations under the
null hypothesis that no (cluster) signal is there, which differs
from how probable a signal can be there without detecting it. A pedagogical
astronomer oriented explication of the difference of the two concepts 
is presented in Andreon (2008).

\section{Mixture modeling for $\sigma_v$ measurements}

\begin{figure}
\psfig{figure=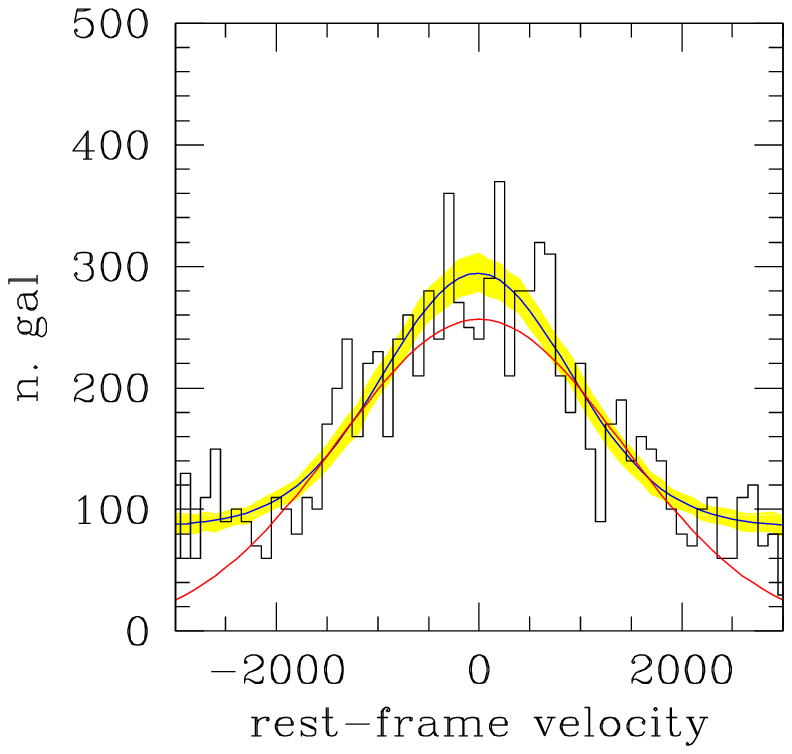,width=8truecm,clip=}
\caption[h]{Velocity histogram for a simulated dataset composed
of 500 member galaxies drawn from a Gaussian having
true $\sigma_v=1000$ km/s and 500 interlopers
drawn from an uniform distribution. The red curve is the
cluster velocity estimate derived from robust statistics
($\widehat{\sigma_v}=1400$ km/s), whereas the blue curve and the
shaded yellow region show the Bayesian estimate: $\sigma_v=940\pm85$ km/s.
See text for details.}
\end{figure}

We want to measure the scale (dispersion) of a distribution (say, of
velocities), knowing that the sample is contaminated by the presence of
interlopers, but without the knowledge of which object is an interloper. The
main idea is not identify or de-weight interlopers in the scale estimate, but
account for them statistically, precisely as astronomers do with photons when
estimating the flux of a source in presence of a background.  The small size of
astronomer samples  (e.g. of cluster galaxies with known velocity) makes the
asymptotic  properties of frequentist estimators never reached in real life
experiments and oblige us to look for a solution in the Bayesian paradigm. 

Here, we assume that data come from two populations: background 
galaxies, whose distribution is assumed to be an homogeneous (i.e. the intensity is
independent on $v$) Poissonian random process, and cluster galaxies, whose distribution
is assumed to be a Poissonian process whose intensity is given by a Gaussian. 
The likelihood is given by eq A1, with changes of
variable names: $\Omega$ continues to be $\Omega$, but it is easier to understand 
it if we call it $\Delta v$, the (velocity) range over which velocities are 
considered (say, $\pm 5000$ km/s from the cluster preliminary velocity center). 
$\Omega$ is more appropriate than $\Delta v$ as it accounts for intervals of complicated 
shape; $r$ in Appendix A is now $v$, and $I(v)$ is given by the sum of a Gaussian and a 
constant, with unknown weights, $N_{clus}$ and $N_{bkg}$ (respectively). Finally, each 
measured velocity $v$ has an uncertainty $\sigma_{v}$. Therefore, $I(v)$ reads:

\begin{equation}
I(v) =  \frac{N_{clus}}{2 \pi \sqrt{\sigma_v^2+\sigma_{clus}^2}} e^{-\frac{(v-v_{clus})^2}{2 
(\sigma_v^2+\sigma_{clus}^2)}} + \frac{N_{bkg}}{\Delta v}
\end{equation}

Most literature estimates of cluster velocity dispersions are based on 
the family of estimators presented by Beers, Flynn \& Gebhardt (1991). However, 
in the presence of a background and interlopers, with sparsely sampled data, a
Bayesian estimator may be more appropriate. 

A `real life' example may suffice. The cluster Cl1604+4304 was regarded as unusually
X-ray dim for the large mass estimated from a sample of 27 redshifts $\widehat{\sigma_v}=
1226^{+245}_{-154}$ km s$^{-1}$ (Postman, Lubin \& Oke 2001). However, the Bayesian
method returns $\sigma_v=1022 \pm 570$ km s$^{-1}$ (posterior mean and
standard deviation), which no longer makes
the cluster X-ray underluminous and has a more realistic error bar. A larger sample
of redshifts for this cluster, from Gal \& Lubin (2004) revises the original estimate
to $\widehat{\sigma_v}=962 \pm 141$ km s$^{-1}$, in good agreement with the Bayesian estimate.
With this value, the cluster is no longer X-ray underluminous.

Let us now consider a simulated 'cluster' composed of 500 galaxies distributed in a
Gaussian with $\sigma_v=1000$ km s$^{-1}$ and superposed over a background of 500
uniformly distributed (in velocity) interlopers. The large sample size has been adopted
to leave data to speak by themselves. Applying the methods of Beers et al. (1991) yields
$\widehat{\sigma_v}=1400$ km s$^{-1}$ which is an excessively large estimate of
$\sigma_v$ (and hence of mass), as also visible in Fig.~B1 by simple inspection (compare
the red curve and the histogram). The Bayesian posterior mean is $\sigma_v=940 \pm 85$
km s$^{-1}$ which is closer to  the `true' value (blue curve with shading). This
simulation shows that the amplitude of bias of the Beers et al. estimator is systematic
(i.e. it is present even for a large sample), and it is actually independent on the
sample size, provided the  relative fraction of cluster and interlopers is kept,
although harder and harder to note as the sample size decreases because the estimator
variance increases and dominates the scatter.

\begin{figure}
\psfig{figure=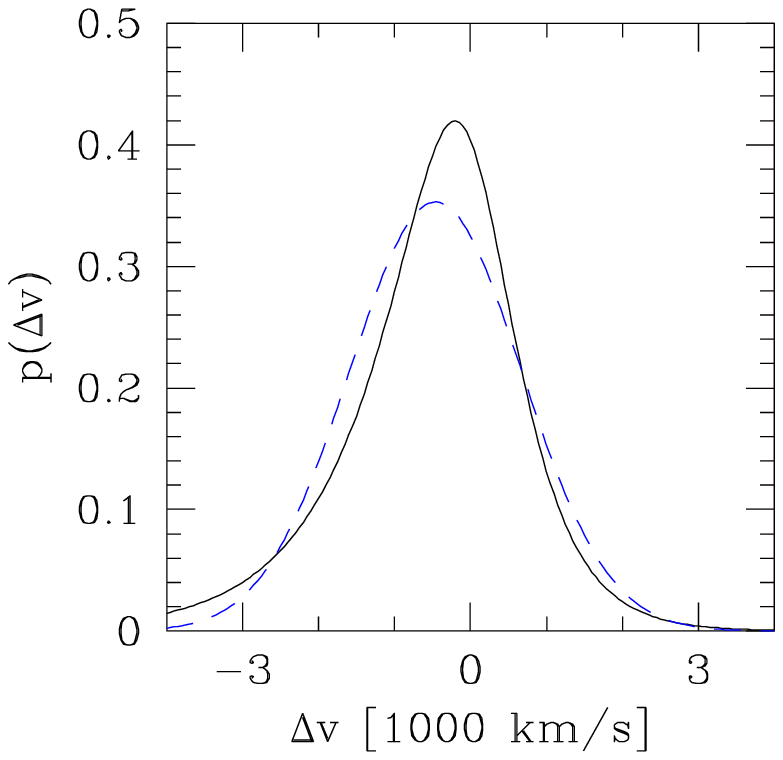,width=8truecm,clip=}
\caption[h]{Perturbed velocity distribution (solid line), given by eq. B2, 
and a Gaussian with identical first two moments (dashed blue line).
The former is used to generate hypothetic data, the latter is assumed to
estimate $\sigma_v$.
}
\end{figure}

We now assess the sensitivity to model assumptions. Lets suppose that
cluster substructure perturbs the velocity distribution, that we now 
assume to be described by 

\begin{equation}
p(v) \propto e^{v/1000} (1+e^{2.75 v/1000})^{-1}
\end{equation}
 
depicted in Fig B2 (solid line). The function has 
first and second moments (mean and dispersion) equal to $-460$ and $1130$
km s$^{-1}$, respectively. We simulate 1000 (virtual) clusters of 25
members each (and no interlopers) drawn from the distribution above (eq B2), 
but we compute the velocity dispersion
using eq. B1, i.e. with a likelihood function appropriate 
for members drawn from a Gaussian.
The mean of found posterior means is $\sigma_v=1140$ km s$^{-1}$ (vs. the
$1130$ km s$^{-1}$ input value) with a standard deviation of $185$ km
s$^{-1}$. The mean error uncertainty (posterior standard deviation) is
$163$ km s$^{-1}$, close (as it should be) to the scatter of the
posterior means. The uncertainty has
a negligible scatter, $18$ km s$^{-1}$,
indicating the low noise level of each individual uncertainty
determination, four time lower than the scatter of the 
uncertainty of the biweight estimator of scale
($70$ km s$^{-1}$), that instead shows 
values as small as $73$ km s$^{-1}$ 
and as large as $865$ km s$^{-1}$ for data that are supposed to
give a unique, fixed, value of uncertainty.


As more difficult situation, we now consider a sample drawn, as before, from a
distribution different from the one used for the analysis, but furthermore
$\sim 50$\% contaminated by interlopers and consisting of half as many members:
13 galaxies are drawn from the distribution above (eq. B2), superposed to
a background of 12 galaxies, uniformly drawn from $\pm5000$ km s$^{-1}$. 
The mean of found posterior means is $\sigma_v=1160$ km s$^{-1}$ (vs. the
$1130$ km s$^{-1}$ input value). 
The mean error uncertainty  is $390$ km
s$^{-1}$, with a low  ($80$ km s$^{-1}$) scatter. 
The biweight estimator returns, on average, a strongly biased
estimate  $\widehat{\sigma_v}=2135$ km s$^{-1}$.



The Bayesian determination of the cluster velocity dispersion 
already embodies the correction for the Eddington 
bias: the prior (i.e. the number distribution
of objects having $\sigma_v$) does matter when the likelihood
is shallow (i.e. when the data does not tightly constraint  the
aimed quantity), because, as well known to astronomers,
if there are many more low velocity systems than high velocity 
dispersion systems the observed value 
(i.e. the maximum likelihood value) is a biased estimate of the 
`true' value. As point out by Jeffreys (1938), 
the Bayes theorem quantifies the bias, and we
used it for computing the correction to
Fang et al. (2007) velocity dispersions. Specifically,
we assume a logarithmic slope of $-0.6$ for the prior
and we follow 
appendix A of Andreon et al. (2006), because
Fang et al. (2007) do not publish individual velocities
for their systems.

Beers, Flynn \& Gebhardt (1991) scale
estimators work correctly in many cases, as shown in their
paper. In these cases, the Bayesian approach returns similar
numbers. We have shown, 
however, that in frontier-line cases,
i.e. in the presence of an 
important background, or with sparsely sampled data, the
Bayesian method returns better behaved quantities, less
biased and less noisy.

\label{lastpage}

\end{document}